\documentclass[aps,prb,10pt,twocolumn,floatfix,letterpaper,nofootinbib,longbibliography]{revtex4-2}

\usepackage{graphicx}
\usepackage[export]{adjustbox}

\usepackage{dcolumn}
\usepackage{bm}
\usepackage{multirow}

\usepackage[normalem]{ulem}

\usepackage{amsmath}
\usepackage{amsthm}
\usepackage{amstext}
\usepackage{amssymb}
\usepackage{mathrsfs}
\usepackage{amsfonts}
\usepackage{amsbsy} 

\usepackage{extarrows}

\usepackage[all,matrix,cmtip]{xy}

\usepackage{color}

\usepackage{hyperref}
\hypersetup{
    colorlinks = true,
    allcolors = blue,
}

\definecolor{red}{rgb}{1,0,0}
\definecolor{blue}{rgb}{0,0,1}
\definecolor{dblue}{rgb}{0,0,0.4}
\definecolor{green}{rgb}{0,1,0}
\definecolor{black}{rgb}{0,0,0}
\definecolor{white}{rgb}{1,1,1}

\definecolor{brn}{rgb}{.8,.4,.0}
\definecolor{redo}{rgb}{1,.5,.0}
\definecolor{ddgrn}{rgb}{0,0.4,0}
\definecolor{dgrn}{rgb}{0,0.55,0}
\definecolor{dbl}{rgb}{0,0,0.5}

\usepackage[bbgreekl]{mathbbol}

\newcommand{\onebb}{\mathbb{1}}

\newcommand{\Z}{\mathbb{Z}}
\newcommand{\C}{\mathbb{C}}

\newcommand{\ii}{\hspace{1pt}\mathrm{i}\hspace{1pt}}
\newcommand{\ee}{\hspace{1pt}\mathrm{e}}
\newcommand{\dd}{\hspace{1pt}\mathrm{d}}
 
\renewcommand{\>}{\rangle} 

\newcommand{\Rf}[1]{Ref.~\cite{#1}}

\newcommand{\prt}{\partial}

\newcommand{\up}{\uparrow} 
\newcommand{\down}{\downarrow} 

\newcommand{\ie}{{\it i.e.~}} 
 
\newcommand{\etc}{{\it etc.}}

\newcommand{\bpm}{\begin{pmatrix}}
\newcommand{\epm}{\end{pmatrix}}
\newcommand{\bmm}{\begin{matrix}}
\newcommand{\emm}{\end{matrix}}

\newcommand{\cC}{ {\cal C} } 
\newcommand{\cD}{ {\cal D} } 
\newcommand{\cE}{ {\cal E} } 
\newcommand{\cF}{ {\cal F} } 
\newcommand{\cG}{ {\cal G} } 
\newcommand{\cH}{ {\cal H} }

\newcommand{\cV}{ {\cal V} }

\usepackage{euscript}

\newcommand\eB           {\EuScript{B}}

\newcommand\eM          {\EuScript{M}}
\newcommand\eN         {\EuScript{N}}

\newcommand\eT         {\EuScript{T}}

\newcommand\eZ         {\EuScript{Z}}

\newcommand\sA           {\mathsf{A}}

\newcommand{\al}{\alpha} 
\newcommand{\bt}{\beta} 
\newcommand{\del}{\delta} 
\newcommand{\Del}{\Delta} 
 
\newcommand{\veps}{\varepsilon} 
\newcommand{\ga}{\gamma}

\newcommand{\la}{\lambda}

\newcommand{\si}{\sigma} 
 
\newcommand{\vphi}{\varphi} 

\newcommand{\frmbox}[1]{\begin{center}\fbox{\parbox{3.3in}{\parindent=0pt #1}}\end{center}}

\usepackage{tikz}
\graphicspath{{figures/}}

\usepackage[a-1b]{pdfx}

\usepackage{booktabs}
\usepackage{microtype}
\usepackage{extarrows}

\begin{document}

\title{Holographic theory of non-invertible gravitational anomalies
}

\author{Xiao-Gang Wen}
\affiliation{Department of Physics, Massachusetts Institute of Technology,
Cambridge, Massachusetts 02139, USA}

\begin{abstract}

We review a many-body formulation of gravitational anomalies as obstructions
to realizing the complete low-energy data by a local {\it bosonic} lattice
quantum system of the same dimension. In this formulation, an anomalous theory
can be realized at the boundary of a gapped {\it bosonic} lattice bulk, whose
topological order gives a one-to-one classification of the anomaly. An
invertible bulk reproduces conventional anomaly inflow; a bulk with
non-trivial topological excitations yields a non-invertible generalization. To
describe the holographic theory of anomalies, we explain the distinction
between uniquely defined topological order and many versions of topological
quantum field theory and discuss the boundary-to-bulk map. As an application
of the holographic theory of anomalies, we describe morphisms between local
quantum theories (such as quantum field theories). These morphisms are built
from a bulk domain wall and an invertible boundary junction, which give rise
to a sandwich picture.  Chiral boson/fermion theory, Laughlin edges, and
boundaries of $\mathbb{Z}_2$ higher gauge theory with bosonic or fermionic
point-like excitations illustrate the different kinds of information about
non-invertible anomalies contained in transport coefficients, superselection
sectors, and condensation data.  In particular, a fermionic theory can be
viewed as a theory with a particular non-invertible gravitational anomaly.
Those examples also reveal that \emph{symmetry $\sim$ gravitational anomaly
$\sim$ topological order in one higher dimension}.

\end{abstract}

\maketitle

\setcounter{tocdepth}{2}
\tableofcontents

\section{Introduction}

An anomaly was originally understood as the failure of a classical symmetry to
survive quantization, \ie an obstruction to quantizing a theory while
preserving the relevant symmetry. For gravity, the symmetry is a spacetime
symmetry: diffeomorphism invariance or local frame invariance. An uncanceled
gravitational anomaly obstructs a consistent coupling to dynamical
gravity~\cite{AGW1984}. Such defined anomaly is a property of the quantization
procedure. But we want to define anomaly as a property of a \emph{local
quantum theory} specified by a Hilbert space and a Hamiltonian? An example of
a local quantum theory is the chiral fermion theory on a one-dimensional ring:
\begin{align}
\label{cferm}
 H &= -\ii \int_0^L \dd x\; \la(x)\prt_x \la(x),
\nonumber\\
\la^\dag(x) &= \la(x)
, \ \
\{\la(x),\la(y)\} = 2\del(x-y) .
\end{align}
The Hamiltonian is built from the \emph{local operators}, $O(x) = \la(x)\prt_x
\la(x)$, which are defined as commuting operators when they have disjoint
support.  The local operators satisfy a periodic boundary condition
$O(x)=O(L)$, and thus the fermion field can have both periodic and
anti-periodic boundary conditions $\la(0)=\pm \la(L)$. The theory has a
well-defined low-energy spectrum that contains two sectors from the two
boundary conditions.

We remark that the formal continuum expression of a quantum field theory alone
may not specify a nonperturbative theory, especially in the presence of
interactions.  We therefore introduce ``local quantum theory'' as a
nonperturbative definition of quantum field theory: \frmbox{A local quantum
theory in $d$ spatial dimensions is a triple $(\cV,H, \sA)$, where the Hilbert
space $\cV$ is an irreducible representation of a local operator algebra
$\sA$, and the Hamiltonian $H$ is a sum or integral of local operators in that
algebra.} This description presupposes a consistent representation and a
self-adjoint Hamiltonian; it is not, by itself, a construction of either.  The
local operator algebra may have different irreducible representations
corresponding to different superselection sectors.  We like to define anomaly
as a property of local quantum theory.

Refs.~\cite{W1313,KW1458} proposed that \frmbox{a local quantum theory $\cC_d$
in $d$-dimensional space is free of a Hamiltonian-type gravitational anomaly
if it has a lattice ultraviolet (UV) completion, \ie it can be realized as the
low-energy effective theory of a local \emph{lattice quantum system} with a
finite number of states per site in the same dimension.}  A parallel
Lagrangian-type definition uses a local bosonic path integral on a spacetime
lattice.

\begin{figure}[tb]
\centering
\includegraphics[scale=0.7]{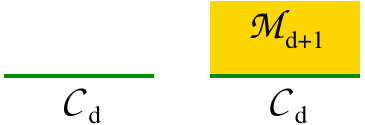}
\caption{$\cC_d$ on the left has a lattice UV completion in the same dimension
and is anomaly-free. On the right, $\cC_d$ is the low-energy boundary theory
of a gapped, bosonic, anomaly-free lattice system $\eM_{d+1}$ in one higher
dimension. Such $\cC_d$ may be anomalous. We regard theories that admit
neither realization as locally inconsistent \cite{KW1458}. We exclude them
from our discussion.  } \label{CM}
\end{figure}

As emphasized in Ref.~\cite{KW1458}, a local quantum theory without a bosonic
lattice UV completion with finitely many states per site in its own dimension
may nevertheless occur at the boundary of a gapped local bosonic lattice
system with finitely many states per site in one higher dimension
(Fig.~\ref{CM}).  Its topological order is the holographic datum proposed to
classify the anomaly in the local quantum theory.

\begin{figure}[tb]
\centering
\includegraphics[scale=0.7]{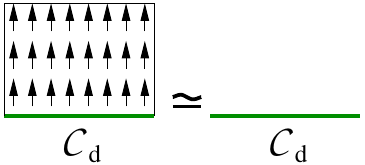}
\caption{$\cC_d$ is a boundary of a trivial topological order (\ie a product
state). The product state can be removed and $\cC_d$ can be realized in the
same dimension (\ie not as a boundary).} \label{CP} 
\end{figure}

This defines the anomaly directly from a local quantum
theory. When the bulk is a product state, it can be removed, leaving a
same-dimensional lattice realization of the boundary theory. The theory is
then free of a gravitational anomaly (Fig.~\ref{CP}). If its anomaly bulk has
nontrivial topological order $\eM_{d+1}$, the bulk cannot be removed while
preserving the complete boundary data.
The proposal is therefore that bulk topological orders classify gravitational
anomalies in one lower dimension~\cite{KW1458}.

Since the bulk topological order need not be invertible, this anomaly need not
be invertible either. We call this extension a \emph{non-invertible
gravitational anomaly}, generalizing conventional anomaly
inflow~\cite{CH8527}, since the extended framework includes invertible
anomalies as a special case.

The framework has three main features:
\begin{itemize}
\item
    \emph{Lattice realizability as a criterion.} A local quantum theory
(whether gapped or gapless) is considered anomaly-free if and only if it can
be UV-completed by a local lattice quantum system with finitely many states
per site (such as a local bosonic lattice Hamiltonian or bosonic
tensor-network path integral) in the same dimension. 

 \item 
    \emph{No spacetime bundles required.} Unlike the traditional high-energy
physics definition, formulated through covariance on smooth manifolds, this
definition uses microscopic locality and quantum information.  The obstruction
concerns a realization of the full theory by a Hilbert space with a
tensor-product decomposition.

\item 
    \emph{Boundary-bulk correspondence.} A local quantum theory with such a
gravitational anomaly cannot exist as an isolated lattice system in its own
dimension. However, it is realized as the boundary of a well-defined gapped
lattice phase in one higher dimension. The claim that every consistent
obstruction admits such a realization is part of the general holographic
proposal.

\end{itemize}

\begin{figure}[tb]
\centering
\includegraphics[scale=0.7]{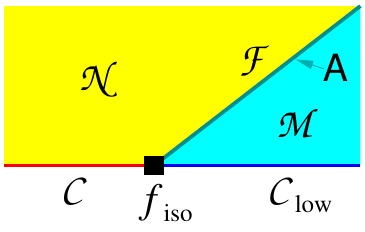}
\caption{ Consider a local quantum theory $\cC$ (such as a quantum field
theory) whose anomaly is described by a bulk topological order $\eN$. An
equivalence $f_\mathrm{iso}$ gives a holographic decomposition of $\cC$ as the
relative stacking of the boundary $\cC_\text{low}$ and the wall $\cF$ over the
shared bulk $\eM$: $\cC\simeq\cF \boxtimes_{\eM}
\cC_\text{low}$~\cite{KZ150201690}. The wall $\cF$ is gapped.  The
decomposition defines a morphism $\cC_\text{low}\to\cC$ and exposes the
anomaly carried by the relative factor $\cC_\text{low}$.  } \label{CCmorph} 
\end{figure}

\begin{figure}[tb]
\centering
\includegraphics[scale=0.7]{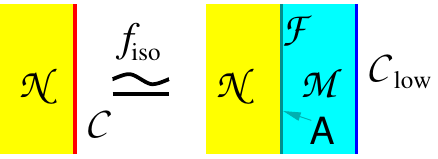}
\caption{A generalized sandwich configuration. The intermediate
bulk $\eM$ lies between the gapped wall $\cF$ and the boundary
$\cC_\text{low}$. When the remaining bulk $\eN$ is trivial, their relative
stacking gives an absolute lower-dimensional theory.}
\label{sandwich}
\end{figure}

The holographic theory of anomalies also suggests a way to organize relations
between local quantum theories. Quantum field theory describes an enormous
range of physical systems, including strongly coupled theories without a
useful Lagrangian or a preferred set of fundamental fields. In mathematics, a
morphism is a structure-preserving map and is a basic tool for understanding
mathematical structures. To better understand quantum field theories,
including those without a Lagrangian or weakly coupled fields, it is therefore
important to define morphisms between them.

Because local quantum theories (such as quantum field theories) may themselves
be anomalous, the holographic picture implies that a morphism between local
quantum theories must retain their anomaly-characterizing bulks.  For gapped
liquid states, this idea is implemented in Definition~4.3 of
Ref.~\cite{KZ150201690}.  Its extension to gapless theories was proposed in
\Rf{KZ170200673,KZ170501087,KZ190504924}.  Both use the configuration in
Fig.~\ref{CCmorph} to define:  
\begin{align}
 \cC_\text{low} \xrightarrow{\text{morphism}} \cC
\end{align}
where $\cC$ and $\cC_\text{low}$ represent gapped or gapless local quantum
theories.

Figure~\ref{CCmorph} generalizes the sandwich picture discussed in
Ref.~\cite{FT220907471}. It can be redrawn as Fig.~
\ref{sandwich}. When the bulk topological order $\eN$ is trivial, the right
part of Fig. \ref{sandwich} becomes a sandwich.

We remark that, in much of the modern literature, ``anomaly'' is reserved for
an invertible anomaly theory. Instead of ``non-invertible anomaly,'' some
papers use ``anomaly topological quantum field theory'' or ``gravitational
anomaly described by a topological quantum field theory.''

Throughout, the terms ``local quantum theory'' and ``local lattice quantum
system'' have the operational meanings given above and in Sec.~\ref{llqs}. We
use $D$ for spacetime dimension and $d$ for spatial dimension.  ``Anomaly'' in
this paper always refers to a gravitational anomaly. We do not discuss
symmetry anomalies, 't Hooft anomalies, or gauge anomalies.

\section{Topological order versus topological quantum field theory}

Topological order and topological quantum field theory (TQFT) both play
central roles in the definition of gravitational anomalies. We introduce these
concepts and explain why they should be distinguished.  We will see that
topological order has an unique definition while TQFT has many different
versions. It is not clear which version of TQFT corresponds to the uniquely
defined topological order.

\subsection{Topological order as a gapped quantum liquid phase on lattice}

A bosonic topological order is a gapped quantum liquid phase of a bosonic local
lattice quantum system with a finite number of states per site.  In this
paper, we discuss only bosonic topological orders and bosonic lattice
systems. We will therefore omit the word ``bosonic'' below. The notion of a ``gapped liquid
phase'' is developed in Refs.~\cite{ZW1490,SM1403} using quantum information.
We explain the relevant definitions below.

\subsubsection{Local lattice quantum systems and gapped quantum systems}
\label{llqs}

\begin{figure}[tb]
\centering
\includegraphics[scale=0.4]{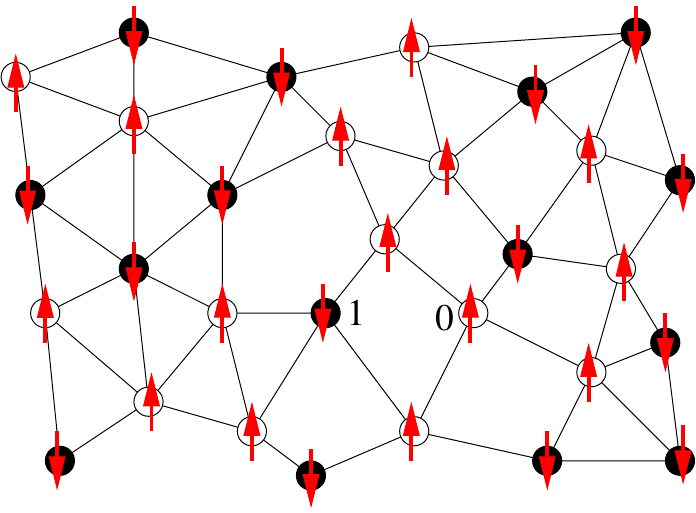}
\caption{A lattice quantum system with $N$ sites. Each site carries a local
Hilbert space $\cV_i$, here taken to be $\C^2$.}
\label{bsnspnspace} 
\end{figure}

A \emph{local lattice quantum system} is described by a triple $(\cV_N,
\otimes, H_N)$, where $\cV_N$ is a Hilbert space, $\otimes$ is a tensor
product decomposition: $\cV_N =\otimes_{i=1}^N \cV_i$, and $H_N$ is a local
Hamiltonian acting on $\cV_N$:
\begin{align}
H_N=\sum_i O_i.
\end{align}
Here $O_i$ acts on $\cV_i$ or on a bounded neighborhood of site $i$.  The
sites form a graph, whose edges define proximity.  We assume uniformly bounded
local interactions with finite range as the
system grows. 

\begin{figure}[tb]
\centering
\includegraphics[scale=0.6]{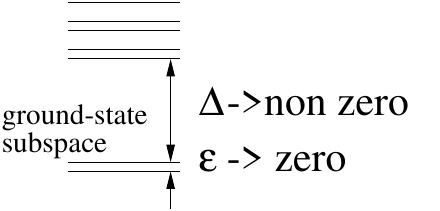}
\caption{The lower end of the eigenvalue spectrum of $H_N$ for a large $N$.
The spectrum is \emph{gapped} if $\Del$ remains bounded away from zero and
the splitting $\veps$ approaches zero as $N\to\infty$. The states below
$\Del$ form the nearly degenerate ground-state subspace.} \label{gap}
\end{figure}

A \emph{gapped quantum system} is a sequence
$\{(\cV_{N_k},\otimes,H_{N_k})\}_{k=1}^{\infty}$ in the thermodynamic limit,
rather than merely a single finite system. For sufficiently large $N_k$, the
low-energy spectrum has the form shown in Fig.~\ref{gap}, with a uniform
gap above the nearly degenerate ground states. These states define a
corresponding sequence of ground-state subspaces.

\subsubsection{Quantum information definition of gapped quantum phases}

Two ground-state subspaces $\cV^\text{grnd}_0$ of $H(0)$ and
$\cV^\text{grnd}_1$ of $H(1)$ are in the same gapped phase if a smooth
path of local Hamiltonians $H(g)=\sum_i O_i(g)$ connects them while
maintaining a uniform gap for $g\in[0,1]$. At finite size, adiabatic
evolution relates the subspaces in the slow-evolution limit:
\begin{align}
\cV^\text{grnd}_1
\simeq
 P\Big( e^{-\ii \int_0^1 d g'\;  \bt H(g')} \Big) \cV^\text{grnd}_0.
\end{align}
Here $P$ denotes ordering along the path and $\bt$ is the evolution time.  For
the thermodynamic limit, quasi-adiabatic continuation gives a more precise
statement: the ground-state subspaces are related by a quasi-local spectral
flow~\cite{HW0541,Bachmann2012}. The adiabatic evolution can be well
approximated by an finite-depth circuit of strictly finite-range gates.

The circuit formulation describes phase equivalence using finite-depth local
unitary (LU) transformations~\cite{ChenGuWen2010}, schematically
\begin{align}
\cV^\text{grnd}_1 \sim
\bmm \includegraphics[scale=0.2]{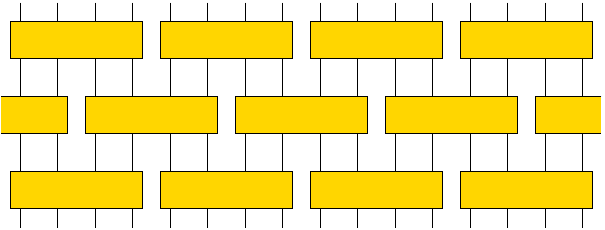} \emm \
\cV^\text{grnd}_0 
\end{align}
Here the circuit depth (\ie the number of layers) remains finite
as the system size $N\to\infty$.

Finite-depth LU transformations define an equivalence
relation on ground-state subspaces. Subspaces related in this way belong to
the same phase; a phase is an equivalence class of subspaces under these
transformations. Comparison with a general gapped Hamiltonian
path requires allowing quasi-local tails or controlled approximations to
local observables, rather than assuming exact equality of finite-size
wave functions.

\subsubsection{Quantum information definition of gapped quantum liquid phases}

\begin{figure}[tb]
\centering
\includegraphics[scale=0.5]{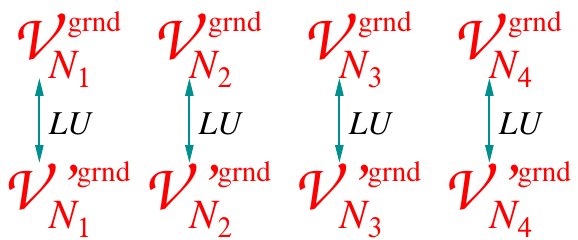}
\caption{LU transformations connect systems with the same number of sites.
} \label{gapLU} 
\end{figure}

\begin{figure}[tb]
\centering
\includegraphics[scale=0.4]{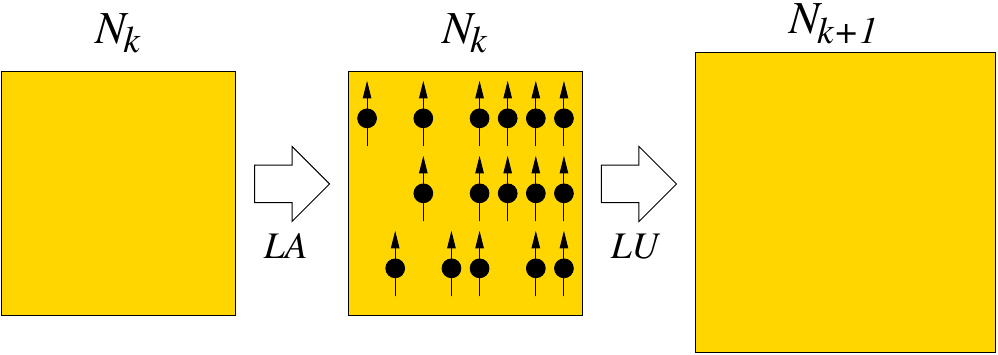}
\caption{
Combining local addition (LA) with finite-depth LU transformations gives
generalized LU transformations.
} \label{NkNk} 
\end{figure}

The fixed-size definition does not compare systems with different numbers
of sites (Fig.~\ref{gapLU}). Translation invariance provides one way to
specify a sequence of larger systems, but a definition of phase should not
depend on that symmetry.

To compare systems without translation invariance, we introduce local
addition (LA): stacking a product state, as in
Fig.~\ref{NkNk}. The enlargement step is
\begin{align}
\widetilde{\cV}^{\text{grnd}}_{N_{k+1}}
\overset{LA}{\sim}\cV^\text{grnd}_{N_k}\otimes
|\text{product state of $N_{k+1}-N_k$ sites}\>.
\end{align}
Combining LA and finite-depth LU transformations
gives generalized LU (gLU) transformations, which connect systems with
different sizes (see Fig. \ref{NkNk}).

\begin{figure}[tb]
\centering
\includegraphics[scale=0.5]{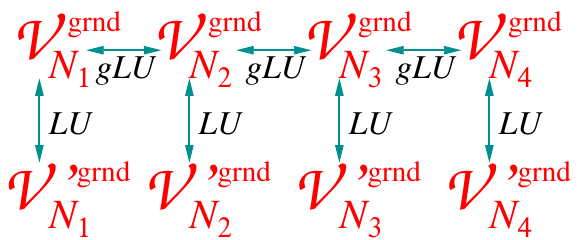}
\caption{Generalized LU transformations connect systems with different
numbers of sites and allow comparison along a thermodynamic sequence
$N_k\to\infty$.} \label{liquidLU}
\end{figure}

The gLU transformations let us compare systems in the thermodynamic limit
without translation symmetry (Fig.~\ref{liquidLU}). A sequence that can absorb
added product-state degrees of freedom through such local transformations has
the liquid property. Equivalence classes of these sequences define
\emph{gapped quantum liquid phases}~\cite{ZW1490,SM1403}.  \frmbox{ A
\emph{topological order}, characterized macroscopically by robust ground-state
degeneracy and non-Abelian geometric phases of degenerate ground
states~\cite{W8987,W9039,WN9077,KW9327}, is defined as a stable gapped quantum
liquid phase \cite{ChenGuWen2010,ZW1490,SM1403}.  }

According to this definition, a state that can be transformed into a
direct-product state through a finite-depth LU transformation has a trivial
topological order. A nontrivial topological order cannot be transformed into a
direct-product state and has a so called pattern of \emph{long-range quantum
entanglement} \cite{ChenGuWen2010}.

\subsection{Mathematical TQFT}

The physical and mathematical formulations of TQFT are different.  A physical
TQFT (in a restrict sense) is a special quantum field theory (with quantum
fields) that is independent of the continuously varying spacetime
metric~\cite{Witten1989TQFT}.  Chern-Simons theories (with zero chiral central
charge) provide concrete physical examples.

An axiomatic TQFT~\cite{Atiyah1988} assigns state spaces to closed spatial
manifolds and linear maps to bordisms, compatibly with gluing.  Different
Chern-Simons theories may give rise to the same axiomatic TQFT.

These axiomatic and field-theoretic formulations developed around the same
time as the notion of topological order~\cite{W8987}. Their shared
terminology reflects a close physical connection.

Neither formulation should be identified without qualification with the above
defined topological order. Distinct field-theory presentations can describe
the same topological order. For example, in $(2+1)$ dimensions, $U(1)_2$
Chern-Simons TQFT and $SU(2)_1$ Chern-Simons TQFT correspond to the same
semion topological order. 

There are many mathematical TQFTs in $(1+1)$ dimensions, whereas stable gapped
bosonic chains without imposed symmetry have only the trivial product-state
phase.  A useful restriction to TQFT is to require a unique vacuum on a closed
spatial sphere $S^d$, which is a reference geometry, not a contractible space.
This condition alone is not sufficient. The TQFTs defined in \Rf{Atiyah1988}
with this condition still do not correspond to topological order.  To compare
axiomatic TQFTs with topological order, one must restrict the axioms by
physical conditions such as unitarity, suitable finiteness, and the
appropriate tangential structure, and often retain the local and defect data
of an extended theory. Even with these restrictions, a general realization and
classification theorem is not known in all dimensions.

We therefore discuss the proposed correspondence
\begin{align}
\label{anotop}
&\ \ \ \ \text{(potentially non-invertible) gravitational anomaly} 
\nonumber\\
&= \text{topological order in one higher dimension},
\end{align}
with its stated assumptions, rather than an unspecified identification
\begin{align}
&\ \ \ \ \text{gravitational anomaly} 
\nonumber\\
&= \text{TQFT in one higher dimension},
\end{align}
because the required version of TQFT that corresponds to bosonic topological
order on lattice must itself be determined.  Gravitational anomaly and
topological order is directly related since both definitions are based on
lattice UV-completion: gravitational anomaly is an obstruction to have a
lattice UV-completion and topological order is a pattern of long range
entanglement on a lattice UV-completion.

\section{The traditional, invertible anomaly}

\subsection{Failure of covariance after quantization}

Let $Z[g]$ be the partition function of a quantum field theory in
$D$-dimensional spacetime in a background metric $g$. Under a diffeomorphism
$\phi$, an anomalous theory may obey
\begin{equation}
\label{eq:phase-anomaly}
 Z[\phi^{*}g]=\mathcal{A}_{\phi}(g)Z[g],
 \qquad \mathcal{A}_{\phi}(g)\in U(1),
\end{equation}
rather than strict invariance. Infinitesimally this is expressed by a failure
of stress-tensor conservation,
\begin{equation}
\label{eq:ward}
 \nabla_{\mu}\langle T^{\mu}{}_{\nu}\rangle
 =\mathcal{G}_{\nu}(g),
\end{equation}
possibly accompanied by a local Lorentz anomaly. In the path integral
language the classical action may be invariant while the regulated measure is
not. Fujikawa's Jacobian method made this mechanism particularly
transparent~\cite{Fujikawa1979}, and Alvarez-Gaum\'e and Witten gave the general
index-theoretic treatment of perturbative gravitational anomalies in chiral
theories~\cite{AGW1984}.

For an even-$D$-dimensional spacetime boundary, the local perturbative anomaly
is summarized by a closed $(D+2)$-form anomaly polynomial $I_{D+2}$. Locally,
\begin{equation}
\label{eq:descent}
 I_{D+2}=\dd I^{(0)}_{D+1},\qquad
 \delta I^{(0)}_{D+1}=\dd I^{(1)}_{D},
\end{equation}
and the variation of the boundary effective action is the descent integral of
$I^{(1)}_D$. Large diffeomorphisms can reveal global anomalies invisible to
the local polynomial. They are naturally expressed through determinant or
Pfaffian line bundles and through the Atiyah-Patodi-Singer $\eta$ invariant
\cite{APS1975,Witten1985,WittenYonekura2019}.

More precisely, $Z[g]$ is not always a number that
changes by a phase. Rather, the anomalous partition function is a section of
a line bundle over background fields modulo gauge transformations.
The anomaly obstructs a trivialization consistent with locality and gluing.

\subsection{Inflow and invertibility}

Anomaly inflow embeds the theory in $D$-dimensional spacetime as the boundary
of a system in $(D+1)$-dimensional spacetime whose variation cancels
Eq.~\eqref{eq:ward}.  Perturbatively the bulk contains the Chern-Simons term
$I^{(0)}_{D+1}$; globally it is described by an exponentiated $\eta$ invariant
or a related bordism invariant. Callan and Harvey's domain-wall fermions gave
an early physical realization of this cancellation between boundary spectral
flow and a bulk current~\cite{CH8527}.

\begin{figure}[tb]
\centering
\includegraphics[scale=0.6]{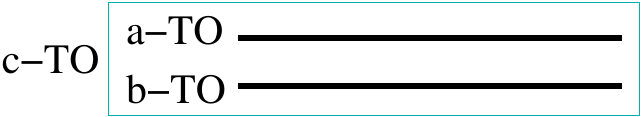}
\caption{Stacking topological orders $a$ and $b$ produces
$c:=a\boxtimes b$. The layers may be coupled by local perturbations that
preserve the gap. Exchanging their order gives an equivalent phase,
$a\boxtimes b\simeq b\boxtimes a$.} \label{stackTO}
\end{figure}

The anomaly theory $\mathcal{I}$ in this standard description is
\emph{invertible}.  Under stacking (see Fig. \ref{stackTO}) there is a
conjugate phase $\overline{\mathcal{I}}$ such that
\begin{equation}
 \mathcal{I}\boxtimes\overline{\mathcal{I}}\simeq \onebb,
\label{eq:inverse}
\end{equation}
where $\onebb$ corresponds to a trivial product state.  On every closed
spatial manifold its state space is one-dimensional, and its closed-spacetime
partition function is a nonzero phase after removal of nonuniversal local
terms. It has no intrinsic topological quasiparticles.  This is why ordinary
anomalies add and admit inverses: their deformation classes form an Abelian
group. The formulation of anomalies as invertible or relative field theories
makes these facts structural rather than accidental
\cite{FT1292,Freed2014}.

This view already treats an anomaly as a property of a self-contained quantum
theory: it is an obstruction to gauging a symmetry. If a symmetry in a quantum
theory can be gauged, then it is anomaly-free; otherwise it is anomalous. For
gravity, the obstruction is especially severe: diffeomorphisms are a gauge
redundancy when the metric is dynamical, so the diffeomorphisms must be gauged
in a quantum theory of gravity.  What is new in the many-body definition is
the microscopic same-dimensional lattice realization test and the possibility
of a non-invertible inflow phase.  Moreover, microscopic lattice systems do
not have manifest continuous diffeomorphism symmetry, so the conventional
field-theory criterion for gravitational anomaly is not directly available at
the cutoff scale.

\section{Historical route to the lattice UV-completion viewpoint}

\subsection{Same-dimensional no-go results}

The Nielsen-Ninomiya theorem is the clearest early precursor. Under its
assumptions, including locality, translation invariance, and the relevant
regularity conditions on the band Hamiltonian, a noninteracting lattice
fermion system has vanishing net chirality: left- and right-handed continuum
species are
paired~\cite{NielsenNinomiya1981a,NielsenNinomiya1981b}.  It is therefore a
microscopic obstruction to realizing a chiral infrared spectrum on a
same-dimensional lattice. Domain-wall fermions evade the naive setup by
placing the chiral theory on a boundary in one higher
dimension~\cite{Kaplan1992}. These results anticipate the UV-obstruction
and boundary-resolution logic.  

However, this free-fermion theorem does not settle the question for
interacting systems, including bosonic systems. 

\subsection{Quantum Hall bulk-edge correspondence}

Quantum Hall physics supplied the decisive interacting many-body intuition. A
chiral edge cannot be detached from its bulk without losing the microscopic
realization of its purely chiral transport. A $U(1)$ gauge anomaly
accounts for charge inflow and the associated
quantized Hall conductance $\si_{xy}$, while
a gravitational anomaly accounts for the chiral central charge
\begin{equation}
\label{eq:chiral-c}
 c_- = c_R-c_L
\end{equation}
and the associated thermal Hall response. The relation between Chern-Simons
bulk theories, chiral edge theories, and anomaly cancellation was developed in
the early quantum Hall literature~\cite{W9125,KaneFisher1997,Cappelli2002}.
It showed in concrete lattice systems that an anomalous low-dimensional theory
can be perfectly meaningful as boundary data.

The conceptual progression is thus:
\begin{align}
\label{eq:history}
&\text{chiral edge state}\\
&\quad\Longrightarrow\ \text{same-dimensional lattice obstruction}
\nonumber \\
&\quad\Longrightarrow\ \text{lattice inflow from one higher dimension}
\nonumber \\
&\quad\Longrightarrow\ \text{bulk topological order as the full anomaly}.
\nonumber 
\end{align}
The first two arrows have substantial earlier antecedents, showing how
field-theory inflow acquires a microscopic interpretation~\cite{W9125}.
The final arrow, elevated to a general classification principle that allows
intrinsically topologically ordered bulks, is the
distinctive claim of Refs.~\cite{W1313,KW1458}.

\subsection{From string-nets to centers}

Levin-Wen string-net models showed that a unitary fusion category
$\mathcal{C}$ gives a lattice topological order in two-dimensional space whose
anyon theory is the Drinfeld center $Z(\mathcal{C})$ of the fusion
category~\cite{LevinWen2005}. Kitaev and Kong treated gapped boundaries and
domain walls explicitly. Boundary conditions are described using module
categories; the excitations on a fixed boundary form a fusion category. A
canonical boundary has the input fusion category $\cC$ as its excitation
theory, up to orientation conventions~\cite{KitaevKong2012}. These results
supplied a concrete categorical model for the slogan that a boundary
determines a unique bulk.  They are direct mathematical and physical
precursors of Refs.~\cite{KW1458,KZ150201690}.

\subsection{Anomalies and relative quantum field theories}
\label{sec:anomalies-relative-qft}

Freed and Teleman formulated a broad generalization of anomaly
inflow~\cite{CH8527} using \emph{relative quantum field
theory}~\cite{FT1292}. An ordinary, or absolute, quantum field theory
${\eT_D}$ in $D$-dimensional spacetime assigns a complex-valued partition function
$Z_{\eT_D}(M^D)\in\mathbb{C}$ to every closed $D$-dimensional manifold $M^D$ and a
Hilbert space $\mathcal{H}_{\eT_D}(N^{D-1})$ to every closed $(D-1)$-manifold
$N^{D-1}$.  

Here ``absolute'' means that the field theory is defined without
an additional anomaly theory. It does not by itself establish a lattice UV
completion with finitely many states per site. The latter is the stronger
microscopic question addressed in Sec. \ref{UVano}.

The absolute theory ${\eT_D}$ may be gapless. Its partition function
$Z_{\eT_D}(M^D)$ and Hilbert space $\mathcal{H}_{\eT_D}(N^{D-1})$ may depend
on the metric and other background data. If $N^{D-1}$ bounds a manifold $M^D$,
a path integral on $M^D$ prepares a state on $N^{D-1}$. State preparation may
require suitable boundary data. 

A relative quantum field theory $\eT_{D-1}^\text{rel}$ in $(D-1)$-dimensional
spacetime $N^{D-1}$, modeling an anomalous theory, is defined only in relation
to an absolute quantum field theory ${\eT_D}$ in $D$-dimensional spacetime
$M^D$, which may be gapped or gapless. Its partition function is not
canonically a complex number, but instead takes values in the state space
assigned by ${\eT_D}$:
\begin{equation}
    Z_{\eT_{D-1}^\mathrm{rel}}(N^{D-1})\in \cH_{\eT_D}(N^{D-1}).
\end{equation}
The higher-dimensional absolute theory $\eT_D$ thus supplies the anomaly
data for the relative theory $\eT_{D-1}^\mathrm{rel}$. Different classes of
absolute theories lead to different notions of anomaly.

When the absolute theory ${\eT_D}$ is gapped and invertible, the vector space
$\cH_{\eT_D}(N^{D-1})$ is one-dimensional.  It is therefore locally isomorphic
to $\mathbb{C}$, but it need not admit a canonical and globally consistent
trivialization. After making a local choice of trivialization, the partition
function may be represented as a complex number, but under a gauge
transformation or diffeomorphism it can acquire a nontrivial phase:
\begin{equation}
    Z_{\eT_{D-1}^\mathrm{rel}}(N^{D-1},A^{g})
    =
    \ee^{\ii \vphi(N^{D-1},A;g)}
    Z_{\eT_{D-1}^\mathrm{rel}}(N^{D-1},A).
\end{equation}
The phase $\vphi$ is the manifestation of the anomaly. Removing the
anomaly requires a coherent trivialization
\begin{equation}
    {\eT_D} \simeq \onebb,
\end{equation}
where $\onebb$ denotes the trivial invertible field theory. If no such
trivialization exists, the relative theory cannot be promoted to an absolute
$(D-1)$-dimensional theory. This covers traditional invertible gravitational
anomalies.

If the absolute theory ${\eT_D}$ is a TQFT, then the vector space
$\cH_{\eT_D}(N^{D-1})$ is finite-dimensional. The partition function of the
corresponding relative theory $\eT_{D-1}^\mathrm{rel}$ is vector-valued with
finitely many components. If the bulk is non-invertible, this extends the
anomaly to the non-invertible case, beyond a one-dimensional anomaly line.
Different choices of TQFT framework retain different data. The central issue
is to identify the version that captures the non-invertible gravitational
anomaly, which is defined as the lattice UV-completion obstruction for a
\emph{local quantum theory}. A local quantum theory need not be a local
lattice quantum system, since the Hilbert space $\cV$ of a local quantum
theory may not have a tensor-product decomposition.

\section{Anomaly as a bosonic lattice UV-completion obstruction}
\label{UVano}

The traditional approach to anomaly describes an obstruction encountered when
passing from a classical to a quantum theory.
We now formulate a gravitational anomaly as a bosonic lattice UV-completion
obstruction, within local quantum theories.

Consider a local quantum theory in $d$-dimensional space. It is characterized
by its low-energy spectrum, fusion and braiding of gapped excitations,
mapping-class-group actions \cite{W9039,KW9327}, and geometric responses,
\etc. Let $\mathcal{C}_d$ describe such a collection of low-energy data.  The
quantum version of the anomaly question is:
\begin{quote}
Does there exist a local lattice quantum system in $d$ spatial dimensions with
a finite number of states per site, whose low-energy theory is $\mathcal{C}_d$,
with no additional low-energy sector?
\end{quote}
If yes, $\mathcal{C}_d$ is anomaly-free. If no, $\mathcal{C}_d$ is anomalous
(see Fig.  \ref{CM}).  In this operational sense, gravitational anomalies are
obstructions to a bosonic lattice UV completion in the same dimension. 

To understand what kind of gravitational anomaly $\mathcal{C}_d$ has, it is
assumed that $\mathcal{C}_d$ occurs as a boundary of a gapped
$(d+1)$-spatial-dimensional local lattice quantum system $\eM_{d+1}$ with a
finite number of states per site \cite{KW1458}.  Then gravitational anomaly
corresponds to the bulk topological order, which leads to a proposed
classification:\frmbox{For the class of boundary-realizable theories, and
assuming the unique-bulk principle, types of gravitational anomalies are in
one-to-one correspondence with topological orders in one higher
dimension~\cite{W1313,KW1458}.} This is a classification of anomalies, not of
boundary theories: many distinct boundaries carry the same anomaly if they
have the same bulk.  One may seek an equivalent formulation using a suitably
specified TQFT, but the necessary framework is not known in full generality.

Several qualifications matter.
\begin{itemize}
\item Here, ``lattice UV completion'' means a local bosonic lattice realization
with a finite number of states per site. 
The lattice need not be periodic and may be irregular.
\item Lattice UV completion does not involve symmetry, which motivates the word
``gravitational.''
\item The definition concerns all universal low-energy data, 
such as the complete low-energy spectrum.
\end{itemize}

We also remark that Ref.~\cite{KW1458} distinguishes two versions of
gravitational anomalies. An H-type theory is anomaly-free if it can be
realized by a ``local'' lattice quantum system in the same dimension with
finitely many states per site. An L-type theory is anomaly-free if it is
realizable by a ``local'' path integral on a spacetime lattice in the same
dimension with finitely many degrees of freedom per site.  The distinction is
useful because a reflection-positive local path integral, a
commuting-projector construction, and a generic Hamiltonian realization are
not automatically equivalent.  Many classification statements in that work are
explicitly conjectural and depend on which version is used.

Consider a possibly gapless local quantum theory $\cC_d$ in $d$ spatial
dimensions that can be realized on a  boundary of a gapped lattice system in
one higher dimension. Let $\eM_{d+1}$ denote its anomaly bulk.  If $\eM_{d+1}$
contains deconfined topological excitations or has a ground-state degeneracy
on some closed spatial manifold, it is non-invertible and cannot obey
Eq.~\eqref{eq:inverse}. This is because under stacking (Fig.~\ref{stackTO}),
ground-state dimensions and independent excitation sectors only multiply.
These properties are stable under sufficiently weak local interlayer
couplings.  Strong interactions can instead drive condensation and change the
phase, but such a transition is not an inverse under gap-preserving stacking.
Therefore, a non-invertible bulk yields a non-invertible lattice UV-completion
obstruction. 

This broader usage of the term ``gravitational anomaly'' should not erase the
conventional one. Anomaly polynomials, $\eta$ invariants, and bordism
classifications capture the invertible part extremely efficiently. The
non-invertible theory asks for additional data: excitations of all
codimensions, their fusion and braiding, and their relations to boundary
excitations.

\section{Boundary determines bulk}

A bulk topological order can have many different boundary phases. Conversely,
the unique-bulk principle assigns a definite anomaly bulk to complete boundary
data. Equation~\eqref{anotop} can thus be expressed as \emph{boundary
determines bulk}. We introduce a boundary-to-bulk map $\eB$ that sends a
boundary phase $\cC_d$ in $d$-dimensional space (which can be gapped or
gapless) to a bulk topological phase $\eM_{d+1}$ in $(d+1)$-dimensional space
(which is always gapped):
\begin{align}
&\eB(\text{boundary phase}) = 
\text{bulk topological order}
\nonumber\\
\text{or } \ \ &
\boxed{
\eB(\cC_d) = 
 \eM_{d+1}
}
\end{align}
The boundary-to-bulk map $\eB$ is functorial because it relates certain
structures in $\cC_d$ and $\eM_{d+1}$. For example, every topological
excitation in $\eM_{d+1}$ is mapped, via a quotient,  to a boundary excitation
in $\cC_d$.  If a bulk excitation is mapped to a trivial boundary excitation,
then we say that the bulk excitation condenses on the boundary.  The
condensing bulk excitations form a condensable algebra  $\sA$ \cite{K13078244}
of the bulk topological order $\eM_{d+1}$ (see Fig. \ref{CCmorph}). The map
$\eB$ is similar to the Drinfeld center $\eZ$ in category theory, but $\eB$
also acts on gapless phases that are not described by higher fusion
categories. Thus the boundary-to-bulk map $\eB$ is more general than the
Drinfeld center $\eZ$.

By definition, the bulk topological order $\eM_{d+1}$ is realized on a
lattice and therefore has no gravitational anomaly: $\eB(\eM_{d+1}) = \onebb$,
where $\onebb$ represents a trivial phase, \ie a product state.  This leads to
an important property of the boundary-to-bulk map: \emph{the bulk of a bulk is trivial}
\begin{align}
 \eB(\eB( \text{-} )) =  \eB^2(\text{-}) = \onebb .
\end{align}
This is dual to the well-known statement that the \emph{boundary of a boundary is empty}.
 
This suggests a cochain-like organization of phases in different dimensions:
\begin{align}
\label{cochain} 
\cdots \xrightarrow{\eB} \mathsf{T}_{d-1}
\xrightarrow{\eB} \mathsf{T}_{d} \xrightarrow{\eB} \mathsf{T}_{d+1}
\xrightarrow{\eB} \cdots. 
\end{align} 
Here $\mathsf{T}_d$ denotes the collection of boundary-realizable phases in
$d$ spatial dimensions, which can be gapped or gapless. Stacking gives a
commutative monoid, not generally an Abelian group, so this is not an ordinary
linear cochain complex.  The image consists of anomaly-free bulk topological
orders.  This organization and its proposed functorial refinement are central
results of Refs.~\cite{KW1458,KZ150201690}. We next discuss their implications
for morphisms.

\section{Morphisms between gapped and gapless systems: the sandwich picture}

We know that, in mathematics, morphism is one of most importance notions used
to build understanding of most mathematics structures.  Therefore, to gain a
deeper understanding of gapped and gapless local quantum theories (which can
be strongly correlated), it is important to define morphism between two local
quantum theories (for example two quantum field theories). 

\subsection{Why an ordinary wall is insufficient}

If local quantum theories, $\cC_\text{low}$ and $\cC$, were anomaly-free, one
might try to define a morphism simply as a codimension-one domain wall between
them:
\begin{align}
\label{wall}
\text{
\begin{tikzpicture}[baseline=(current bounding box.center)]
  \draw[thick] (0,0) -- (2,0) node[midway, below=2pt] {$\cC$};
  \node[draw, fill=black, inner sep=3pt] at (2,0) {};
  \node at (2,-0.4) {$m$};
  \draw[thick] (2,0) -- (4,0) node[midway, below=2pt] {$\cC_\text{low}$};
\end{tikzpicture}
}
\end{align}
But an anomalous theory cannot exist as a complete stand-alone system.
A proposed wall must also explain how their anomalies (\ie the bulk
topological orders) are related. The naive picture \eqref{wall} suppresses
essential data.  

\begin{figure}[tb]
\centering
\includegraphics[scale=0.7]{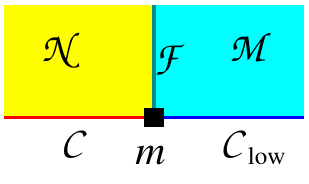}
\caption{ A domain wall between two potentially anomalous systems $\cC$ and
$\cC_\text{low}$. The two boundaries determine the bulks:
$\eN=\eB(\cC)$ and $\eM=\eB(\cC_\text{low})$.
Under the unique-bulk hypothesis, the junction $m$,
together with its neighboring phases, determines its bulk wall $\cF$;
schematically, $\eB(m)=\cF$.}
\label{CCwall} 
\end{figure}

One therefore includes a domain wall extending into the bulk, as in
Fig.~\ref{CCwall}.  But such a definition is not vary useful, since any two
systems can a have a domain wall between them.  We can make the definition
more useful by requiring the domain wall to be gapped.  Such defined morphism
will lead to Witt equivalence classes in the appropriate
categorical setting. 

Definition~4.3 of Ref.~\cite{KZ150201690} provides a different refinement by
requiring $m$ to be an isomorphism, leading to Fig.~\ref{CCmorph}.  It defines
a morphism $f:\cC_\text{low} \to\cC$ as a pair
\begin{equation}
\label{eq:morphism}
 f=\bigl(\cF,f_\mathrm{iso} \bigr),
\end{equation}
with the following meaning (where all dimensions are spatial dimensions):
\begin{enumerate}
\item $\cF$ is a $d$-dimensional gapped domain wall between the
$(d+1)$-dimensional bulks
$\eN=\eB(\cC)$ and $\eM=\eB(\cC_\text{low})$.
\item At the place where that bulk wall meets the boundary, there is an
invertible $(d-1)$-dimensional wall
\begin{equation}
\label{eq:corner}
 f_\mathrm{iso}:
 \cF \boxtimes_{\eM} \cC_\text{low}
 \ \simeq\ \cC.
\end{equation}
Here $\boxtimes_{\eM}$ is a stacking, or a relative tensor
product, over the common bulk $\eM$.
\end{enumerate}

Thus the morphism is not merely a domain wall. It is a layered
bulk-wall-boundary configuration. The first component matches the possibly
non-invertible anomalies; the second asserts that, after fusion with the
source boundary, no residual phase remains at the junction beyond the target
boundary.

\subsection{Why the corner wall must be invertible}

\begin{figure}[tb]
\centering
\includegraphics[scale=0.7]{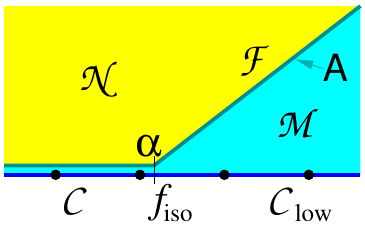}
\caption{ A realization of the isomorphism $f_\mathrm{iso}$ (an invertible
domain wall) in Fig. \ref{CCmorph}.  The vertical line at $f_\text{iso}$
represents an automorphism $\al$ of the structure $\cF \boxtimes_{\eM}
\cC_\text{low}$.  } \label{CCmorphF} 
\end{figure}

Invertibility makes Eq.~\eqref{eq:corner} an equivalence between the fused
configuration and the target theory. A non-invertible junction
can also be meaningful, but defines a more general defect with additional
data, rather than a morphism of this particular kind. The condition ensures
that composing the comparison wall with its inverse leaves a transparent
junction, without a residual junction sector.

The boundary-determines-bulk principle applies to a junction together with its
surrounding phases. Thus $f_\mathrm{iso}$, together with its neighboring phases,
determines $\cF$ in Fig. \ref{CCmorph}. When $f_\mathrm{iso}$ is
invertible, $\cF$ also determines $f_\mathrm{iso}$ up to an automorphism $\al$
of the structure $\cF \boxtimes_{\eM} \cC_\text{low}$, as shown in Fig.
\ref{CCmorphF}. The automorphism $\al$ may permute the topological
excitations on the boundary $\cF$ and in the bulk $\eM$, as well as those on
the boundary $\cC_\text{low}$. Figure~\ref{CCmorphF}
illustrates how an invertible comparison wall $f_\mathrm{iso}$ can be
represented as an equivalent geometric presentation, and we can represent the
morphism \eqref{eq:morphism} as
\begin{align}
\label{eq:morphismA}
 f = \big[\cF, \al\big] .
\end{align}

Figure~\ref{CCmorphF} shows that the excitations in
$\cC_\text{low}$ also appear in $\cC$. This gives the following physical
picture: $\cC$ contains a low-energy sector described by
$\cC_\text{low}$, while other high-energy excitations are
described by the gapped bulk $\eM$ and gapped boundary $\cF$.

Thus we may regard $\eN = \eZ(\cC)$ as the gravitational anomaly of the full
theory $\cC$, and $\eM = \eZ(\cC_\text{low})$ as the gravitational anomaly
of its low-energy sector $\cC_\text{low}$. This gives a notion of an
\emph{emergent gravitational anomaly}: $\eM$ is a gravitational anomaly that
appears at low energies.

\begin{figure}[tb]
\centering
\includegraphics[scale=0.7]{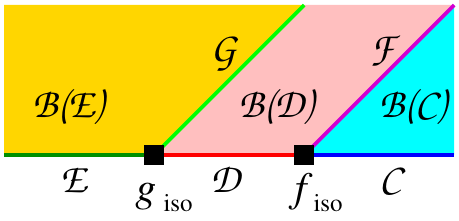}
\caption{Composition of morphisms $f$ and $g$ is described by geometric
composition of transparent domain walls $f_\text{iso}$ and $g_\text{iso}$ on
the lower boundary, together with their bulk structures.} \label{CDEmorph} 
\end{figure}

\begin{figure}[tb]
\centering
\includegraphics[scale=0.7]{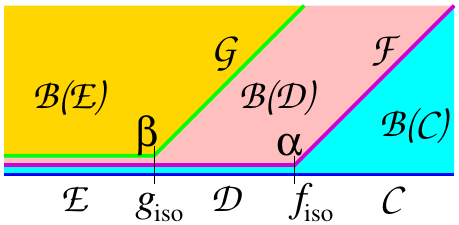}
\caption{ Composition of morphisms $f$ and $g$ is described by geometric
composition of bulk domain walls $\cF$ and $\cG$.  } \label{CDEmorphFG} 
\end{figure}

Figure~\ref{CCmorphF} and the new representation \eqref{eq:morphismA} also show
that morphisms defined in \eqref{eq:morphism} can be composed through
the composition of the bulk part $\cF$ and the automorphism $\al$ even when
the lower boundary $\cC$ and $\cC_\text{low}$ are gapless.  For
$\mathcal{C}\xrightarrow{f}\mathcal{D}\xrightarrow{g}\mathcal{E}$, with
$f=(\cF,f_\mathrm{iso})=[\cF, \al]$ and $g=(\cG,g_\mathrm{iso})=[\cG, \bt]$, we have (see Fig.
\ref{CDEmorph})
\begin{equation}
 g\circ f=\bigl(\cG\boxtimes_{\eB(\cD)}\cF,
 h_\mathrm{iso}\bigr) = \big[ \cG\boxtimes_{\eB(\cD)}\cF, \ga \big],
\label{eq:composition}
\end{equation}
where, with the canonical associativity identifications understood,
\begin{equation}
\label{hgf}
 h_\mathrm{iso}=g_\mathrm{iso}\circ
 \bigl(\mathrm{id}_{\cG}\boxtimes_{\eB(\cD)}f_\mathrm{iso}\bigr).
\end{equation}
One first uses $f_\mathrm{iso}$ to replace $\cF\boxtimes_{\eB(\cC)}\cC$ by
$\cD$, and then uses $g_\mathrm{iso}$ to reach $\cE$. The corner maps, as well
as the bulk walls, enter the composite. 
One can also view \eqref{hgf} as a composition of automorphisms
\begin{align}
\ga = \bt  \circ  \bigl(\mathrm{id}_{\cG}\boxtimes_{\eB(\cD)} \al \bigr),
\end{align}
where $\mathrm{id}_\cG$ is the identity automorphism of
the wall $\cG$.  Associativity is inherited from
wall fusion, with the appropriate coherent identifications.  Thus an abstract
composition law has a local geometric interpretation (Fig.~\ref{CDEmorphFG})
\cite{KZ150201690,KZ170200673}.

\section{Examples of gravitational anomalies}
\label{sec:examples}

The following examples separate three questions: whether the edge has a
chiral energy current, whether the bulk has intrinsic topological sectors,
and whether a chosen boundary can be gapped. These questions are related,
but their answers need not coincide. 

\subsection{Gravitational anomaly in a chiral boson model}

Consider a chiral $(1+1)$D boson model 
\begin{align}
 S_K=\int\dd t\,\dd x\,
 \left[\frac{K_{IJ}}{4\pi}\partial_t\phi_I\,\partial_x\phi_J
       -V_{IJ}\partial_x\phi_I \phi_J\right],
 \label{Kedge}
\end{align}
where $K$ is an invertible symmetric integer matrix with even diagonal entries,
and $\phi_I$ are compact bosonic fields, $\phi_I\sim\phi_I+2\pi$. The theory has
a perturbative gravitational anomaly proportional to
$c_- = \text{signature}(K)$, plus an additional non-invertible gravitational
anomaly if $|\det(K)|\neq 1$.

Such a gravitational anomaly is described by a (2+1)-dimensional
Abelian Chern-Simons theory \cite{Wen1995}:
\begin{equation}
 S_{\mathrm{CS}}=\frac{1}{4\pi}\int K_{IJ}a_I\wedge\dd a_J.
 \label{eq:Kaction}
\end{equation}
Here, the $a_I$ are compact $U(1)$ gauge fields. Quasiparticles are labeled by
integer vectors modulo $K\mathbb{Z}^{r}$, where $r$ is the rank of $K$.
Consequently, there are $|\det K|$ sectors. 

Taking $K$ to be the positive $E_8$ Cartan matrix gives an even unimodular
lattice: $\det K=1$ and $\operatorname{signature}(K)=8$. There are no
fractional quasiparticles, but the edge has eight units of chiral central
charge. Replacing $K$ by $-K$ reverses its chirality and supplies the inverse
phase. Thus an $E_8$ $K$-matrix gives rise to an invertible gravitational
anomaly. A detailed discussion of such bosonic chiral phases is given in
Ref.~\cite{Kitaev2006,Cano2014,Plamadeala2013}.

This example shows why chiral central charge is naturally an invertible
invariant.  We can cancel it by stacking with an opposite-chirality layer. The
thermal Hall conductance is
\begin{equation}
\label{eq:thermalhall}
 \frac{\kappa_{xy}}{T}
 =\frac{\pi^2 k_B^2}{3h}\,c_- , 
\end{equation}
up to the orientation convention.

Equation~\eqref{eq:thermalhall} applies to the universal low-temperature
response, with the edge modes thermally equilibrated.  For a chiral conformal
channel, the equilibrium energy current scales as $j_E=(\pi^2k_B^2/6h)c_-T^2$.
Differentiating the current difference between reservoirs gives the thermal
conductance. Counterpropagating channels contribute with opposite signs. In an
experiment, incomplete equilibration or additional heat carriers can obstruct
a direct extraction of $c_-$~\cite{KaneFisher1997,Cappelli2002}.

Although $c_-$ is additive under stacking, it need not describe a separately
detachable invertible factor. For a bosonic modular anyon theory, the
Gauss-Milgram relation is
\begin{equation}
 \frac{1}{\mathcal{D}}\sum_a d_a^2\theta_a
   =\exp\!\left(\frac{2\pi\ii c_-}{8}\right),
 \qquad \mathcal{D}^2=\sum_a d_a^2,
 \label{eq:gaussmilgram}
\end{equation}
where $d_a$ and $\theta_a$ are the quantum dimensions and topological spins.
Thus the anyon theory determines $c_-$ only modulo eight; an $E_8$ layer
changes the absolute thermal response without changing the anyons.  This is
why the full gravitational anomaly data must include its invertible response
as well as its excitation category~\cite{KW1458}.

\subsection{A fractional quantum Hall edge}

When the $K$-matrix is the one-by-one matrix $K=(m)$, where $m$ is a nonzero
even integer, 
the chiral boson theory 
\begin{equation}
 S_{\partial}=\frac{m}{4\pi}\int\dd t\,\dd x\,
 \left[\partial_t\phi\,\partial_x\phi
       -u(\partial_x\phi)^2\right],\ \ \ u>0,
 \label{eq:laughlin-edge}
\end{equation}
has a gravitational anomaly, described by a topological order that corresponds
to the bosonic Laughlin state 
\begin{align}
 \Psi(\{z_i\})
=
\prod_{i<j}(z_i-z_j)^m \ee^{-\sum_i \frac{|z_i|^2}{4} }, \ \ 
z_i=x_i+\ii y_i .
\end{align}

The Laughlin state has $|m|$ anyon types and $|m|$ ground states on a spatial
torus. It is non-invertible. Its chiral edge has $c_-=\operatorname{sgn}(m)$
and, when charge conservation is retained, a $U(1)$ anomaly is determined by
the Hall conductance $\frac 1m \frac{e^2}{h}$. The same chiral central charge
occurs for many inequivalent fractionalized bulks.

The velocity $u$ is nonuniversal in \eqref{eq:laughlin-edge}, but the allowed
vertex operators $e^{\ii l\phi}$ retain the bulk sector label
$l\in\mathbb{Z}_m$.  Fusion adds these labels modulo $m$. With microscopic
electric charge unit $e$, the bulk anyons have the following quantum numbers
\begin{align}
 \label{eq:laughlin-data}
 q_l&=e\frac{l}{m}\pmod{e}, &
 \theta_l&=e^{\ii\pi l^2/m},\\
 \theta_{ll'}&=e^{2\pi\ii ll'/m}, &
 \sigma_{xy}&=\frac{e^2}{h}\frac{1}{m}.
\end{align}
Here $q_l$ is the fractional electric charge, $\theta_{ll'}$ is the phase for
taking one anyon around another. On a genus-$g$ surface the ground-state
degeneracy is $m^g$. 

Consequently, the conventional gravitational anomaly coefficient $c_-$ sees
only a projection of the holographic anomaly. The full non-invertible bulk
also records topological spin, modular $S$ and $T$ data, and ground-state
degeneracy. No anomaly polynomial can reconstruct this information. This is
the simplest reason that the generalized gravitational anomaly must be
category-valued rather than merely number-valued.

Fractional charge is additional symmetry-enriched information: it requires
specifying the $U(1)$ symmetry and its coupling. The generalized gravitational
anomaly without internal symmetry still distinguishes the intrinsic fusion and
braiding data after charge labels are forgotten.

The spacetime-torus partition functions provide a second view. Let $Z_l(\tau)$
be the partition function of the chiral boson theory \eqref{eq:laughlin-edge}
in the sector $l$ created by $\ee^{\ii l \phi}$, where $\tau$ is the
spacetime-torus modulus. This vector-valued partition function satisfies
\begin{align}
\label{modtrn}
 Z_l(-1&/\tau)=\sum_{l'=0}^{m-1}S_{ll'}Z_{l'}(\tau), \ \ 
 Z_l(\tau+1)=\sum_{l'=0}^{m-1}T_{ll'}Z_{l'}(\tau),
\nonumber\\
 S_{ll'} &= \frac{e^{-2\pi\ii ll'/m}}{\sqrt m}, \ \ 
 T_{ll'}=\delta_{ll'}e^{2\pi\ii(l^2/2m-1/24)}.
\end{align}
A large diffeomorphism therefore acts on the vector-valued partition function
through unitary matrices, not merely through a phase. The factor involving
$1/24$ records the chiral response, while the $S$ and $T$ matrices record the
modular data of the bulk topological order~\cite{JiWen2019}. The modular
transformation property \eqref{modtrn} of the partition function for the
chiral boson theory \eqref{eq:laughlin-edge} is important. It shows how a
non-invertible gravitational anomaly can constrain or even determine the
dynamics of an anomalous theory.

Even an anomaly-free theory $\cC$ may have a hidden gravitational anomaly, as
shown by the holographic isomorphic decomposition in Fig. \ref{CCmorph}.
The gravitational anomaly in a low-energy sector $\cC_\text{low}$ of the
anomaly-free theory $\cC$ can help us determine the low-energy dynamics of
$\cC$. This is the most important application of an emergent non-invertible
gravitational anomaly.

\subsection{A gapped boundary and the Drinfeld center}

Let $\mathcal{C}$ be a unitary fusion category used as input to a $(2+1)$D
Levin-Wen model \cite{LevinWen2005}. The bulk excitations form the braided
fusion category given by the Drinfeld center of $\cC$: $\eZ(\mathcal{C})$.
The bulk topological order $\eZ(\mathcal{C})$ has a canonical gapped boundary.
The gapped topological excitations on the boundary form the fusion category
$\mathcal{C}$.  Thus, a gapped theory in (1+1)D has a gravitational anomaly
$\eZ(\mathcal{C})$ if its topological excitations are described by the fusion
category $\cC$.

For $\mathcal{C}=\mathrm{Vec}_{\mathbb{Z}_2}$, the bulk is the $\Z_2$
topological order \cite{RS9068,W9164} realized by the toric code model
\cite{K032}. The $\Z_2$ topological order has four anyon types:
$1,e,m,f$. A boundary of the $\Z_2$ topological order is induced by $m$
condensation (\ie by the condensable algebra $\sA=\onebb\oplus m$). Such a
boundary has a conserved, gapped $\Z_2$ excitation $e$, in addition to the
trivial excitation $\onebb$. In fact, the boundary corresponds to a
$\Z_2$-symmetric Ising model in the symmetric phase
\begin{align}
 H = -\sum_i X_i X_{i+1} - h Z_i, \ \ \ \ h> 1,
\end{align}
restricted to the symmetric sub-Hilbert space (which is called the symmetric
sector)
\begin{align}
 \cV_\text{sym} = \{|\Psi\> \mid \prod_i Z_i |\Psi\> = |\Psi\> \}.
\end{align}
The ground state is given by $|\Psi_\text{grnd}\> = |\cdots \up\up\up
\cdots\>$ for large $h$, and the excitation $e$ corresponds to a spin $\down$.

The above result can be generalized. A symmetric model on a $d$-dimensional
lattice, with symmetry described by a finite group $G$, becomes anomalous when
it is restricted to its symmetric sector, since the symmetric sub-Hilbert
space $\cV_\text{sym}$ does not have a tensor product decomposition.  Its
gravitational anomaly is described by the topological order of a $G$-gauge
theory in one higher dimension.  Thus, an anomaly-free $G$-symmetric lattice
model has a hidden gravitational anomaly.

This example reveals a general connection: \frmbox{every finite internal
symmetry (including anomalous, higher-group, and non-invertible symmetries)
corresponds to a unique gravitational anomaly \cite{JiWen2019}.} Thus, the
holographic theory of gravitational anomalies reviewed in this article is also
a holographic theory of generalized symmetry: \frmbox{every finite internal
symmetry is described by a topological order in one higher dimension
\cite{KZ200514178}.}

Let us consider a fermionic product state on a $d$-dimensional lattice
\begin{align}
 H = \sum_i c_i^\dag c_i .
\end{align}
When restricted to the even-fermion sub-Hilbert space, with a point-like
fermion as its only nontrivial excitation, the fermionic product state is
described by the symmetric fusion $d$-category,
$d$-$\mathrm{s}{\cV}\mathrm{ec}$ (\ie over a super vector space). The
fermionic product state has a gravitational anomaly because the even-fermion
sub-Hilbert space does not have a tensor-product decomposition.  The anomaly
is described by a topological order in one higher dimension, which is a
twisted higher gauge theory in $(d+2)$-dimensional spacetime $M^{d+2}$.  The
twisted higher gauge theory is described by the following path integral
\cite{GW1441,GK150505856,W161201418,LW180901112}
\begin{align}
\label{Sp2hgauge}
Z = \sum_{\dd f_d=0} \ee^{\ii \pi \int_{M^{d+2}} Sq^2(f_d)}
\end{align}
where $f_d$ is a $\Z_2$-valued $d$-cocycle field. Thus, even a fermion product
state is anomalous: \frmbox{A fermion model on a $d$-dimensional spatial
lattice has a non-invertible gravitational anomaly described by the twisted
higher gauge theory \eqref{Sp2hgauge} in $d+2$-dimensional spacetime.  } As a
result, the physical behavior of a fermion model on a $d$-dimensional spatial
lattice can be simulated on the boundary of the twisted higher gauge theory
\eqref{Sp2hgauge}.

\subsection{The gravitational anomaly in 1+1D chiral Majorana fermion theory}

We have used the 1+1D chiral Majorana fermion theory \eqref{cferm} as an
example of a local quantum theory. What is the gravitational anomaly of such a
local quantum theory?

We can construct the 2+1D bosonic topological order that describes this
anomaly, by starting with the 2+1D $Z_2$ topological order
\cite{RS9068,W9164}.  The $Z_2$ topological order contains a Majorana fermion.
We then make such Majorana fermions to form a spinless $p+\ii p$ topological
superconducting state. The resulting topological order describes the
gravitational anomaly of the 1+1D chiral Majorana fermion theory
\eqref{cferm}.

\section{Conclusion}

The traditional and generalized notions of gravitational anomalies answer
different but compatible questions. The traditional notion detects a failure
of classical symmetry after quantization and is encoded by an invertible
inflow theory.  The generalized notion tests whether complete low-energy data
admit a local same-dimensional lattice realization.  Its inflow bulk can
possess non-trivial topological order in a bosonic lattice model and hence be
non-invertible.

Historically, the lattice obstruction for chiral fermions, anomaly inflow, and
quantum Hall bulk-edge correspondence already tied chirality to an extra
dimension. Refs.~\cite{W1313,KW1458} made the broader proposal that
topological orders in one higher dimension classify generalized gravitational
anomalies. Ref.~\cite{KZ150201690,KZ170200673} then supplied a categorical
mechanism, subject to the unique-bulk hypothesis: the bulk is the appropriate
center, and a morphism between anomalous boundaries is a bulk wall plus an
invertible corner wall. This nested configuration is the core of the sandwich
picture.

The main lesson is that a non-invertible gravitational anomaly is not merely
an anomalous phase factor multiplying the partition function $Z$. It is an
entire topological environment required for the lower-dimensional theory to
exist.  Holography, in this sense, is the statement that the missing
microscopic consistency of the boundary is restored by its associated
higher-dimensional bulk, with functorial organization in the settings where
the wall construction is established.

For a physicist, the examples offer complementary guides.  Chiral energy
transport measures a gravitational response; Laughlin sectors show the
additional information in a vector-valued partition function; and a $\Z_2$
topological order boundary shows that a nontrivial relative theory can be
fully gapped and represent a theory with $\Z_2$ symmetry. These examples lead
to a new understanding of symmetry, including anomalous, higher-group, and
non-invertible symmetries:
\begin{align}
\text{symmetry } &\sim \text{ gravitational anomaly } 
\\
&\sim
\text{ topological order in one higher dimension}
\nonumber 
\end{align}

\bibliography{gravano}

\end{document}